\documentclass{elsart} 

            \usepackage{graphicx} 
            \usepackage{amssymb} 

            \newcommand{\comment}[1]{}

\begin{document} 

            \begin{frontmatter} 


            \title{Elliptic Flow from Color Glass Condensate} 

            \author{Alex Krasnitz$^{\rm a}$, Yasushi Nara$^{\rm b}$, Raju 
            Venugopalan$^{\rm b,c}$} 
            \address{$^{\rm a}$FCT and CENTRA, Universidade do Algarve,  
               P-8000 Faro, Portugal} 
            \address{%
             $^{\rm b}$RIKEN BNL Research Center, Brookhaven National 
            Laboratory, 
                            Upton, N.Y. 11973, U.S.A. 
            } 
            \address{%
             $^{\rm c}$Physics Department, Brookhaven National Laboratory, 
            Upton, N.Y. 11973, 
            U.S.A.   
            } 


            \begin{abstract} 
            We show that an observable fraction of the measured elliptic flow 
            may 
            originate in classical gluon fields at the initial stage of a 
            peripheral 
            high-energy nuclear collision. This mechanism complements the 
            contribution of 
            late stage mechanisms, such as those described by hydrodynamics, 
            to the observed elliptic flow. 

            \end{abstract} 


            \end{frontmatter} 

            The elliptic flow $v_2$, both integral and differential, 
            is a sensitive measure of collectivity of the excited and dense 
            matter 
            produced in ultra-relativistic heavy ion 
            collisions~\cite{Ollitraut}. 
            The first measurements of $v_2$ from RHIC, at center of mass
            energies of 130 and 200 GeV, have been reported 
            recently~\cite{STARv2}. Hydrodynamic (HD) analysis, based on the 
assumption of local thermal equilibrium, matches 
            the data for the integral $v_2$ at large centralities, but the 
agreement gets worse for peripheral events~\cite{KHTS,Hirano:2001eu}.  
            HD models also reproduce 
            the differential $v_2$ up to momenta of 1.5 GeV/c at 
            mid-rapidity. However, 
            above $1.5$ GeV, the experimental $v_2$
            appears to saturate, while the HD model $v_2$ still 
grows~\cite{KHTS}. 
            
            It is natural to expect $v_2$ to be sensitive to the 
            early evolution of the system~\cite{Sorge}, when the 
            energy 
            density of the produced matter is at its highest, and before the 
system has equilibrated. 
            Here we compute the contribution to $v_2$ at mid-rapidity 
            from the strong fields generated shortly 
            after the collision. These fields originate in a 
Color Glass Condensate (CGC)~\cite{MV}, formed in a nucleus by low-x partons 
as their distributions saturate~\cite{GLRMQ}. The CGC is characterized by the 
color charge 
            per unit area $\Lambda_s$ which grows with 
            energy, centrality and the size of the nuclei. 
            Estimates for RHIC give $\Lambda_s\sim 1.4-2$ GeV.  
            Since the gluon multiplicities in CGC are large, 
$\sim 1/\alpha_S(\Lambda_s^2)>1$, CGC admits a classical description. In a 
collision, gluon production results from overlapping CGCs of the incident 
nuclei~\cite{KMW}. Our numerical 
            work~\cite{AR99,AR01} 
            confirmed that strong color fields of order 
            $1/\alpha_s$ emerge in a proper time $\tau \sim 1/\Lambda_s$ 
            after the collision. 

            As before, we assume strict boost invariance, {\it i.e.,} the
dimensionality of the problem is 2+1.
            For a numerical solution we use lattice discretization. 
            Our original setup, suitable for central collisions of very 
            large nuclei, 
            must be adapted for the task at hand. To study effects of 
            anisotropy and 
            inhomogeneity, we consider finite nuclei. We 
            also impose 
            suitable neutrality conditions on the color 
            sources~\cite{Lam:1999} to prevent gluon production far 
            outside the nucleus. 

            We model a nucleus as a sphere of radius $R$, 
            filled with randomly distributed nucleons.
Within each nucleon
we first generate, throughout the transverse plane, a 
            spatially 
            uncorrelated Gaussian color charge distribution 
            of the width $\Lambda_{\rm n}$.  
            Next, we remove the monopole and dipole components of the 
            distribution by 
            subtracting the appropriate uniform densities.
            Since the color charges of the
            nucleons are uncorrelated, $\Lambda_s$ becomes position-dependent,
peaking at the center and vanishing at the periphery of a nucleus. 
            We adjust $\Lambda_{\rm n}$ to ensure a desired value 
            of $\Lambda_{s0}$, {\it i.e.,} $\Lambda_{s}$ at the center.
            Next, we use our standard methods~\cite{AR99}
and determine the classical fields as a function of $\tau$.

            The calculation of $v_2$ 
            involves 
            determining the gluon number $N$, a quantity whose meaning 
            is ambiguous outside a free theory. 
We resolve 
            this 
            ambiguity by computing the number in two different ways; 
            directly in Coulomb Gauge (CG) 
            and by solving a system of relaxation (cooling) equations for the 
            fields~\cite{AR01}. 
            Both definitions give the usual particle number in a 
            free theory. 
            We expect the two to be in good agreement for a weakly coupled 
            theory. 
            If the two disagree strongly, we should not trust either. 
Details of the cooling method, as applied to $v_2$, 
are presented in our recent paper~\cite{AYRv2}.

\begin{figure} 
\begin{center}
            \includegraphics[width=3.0in]{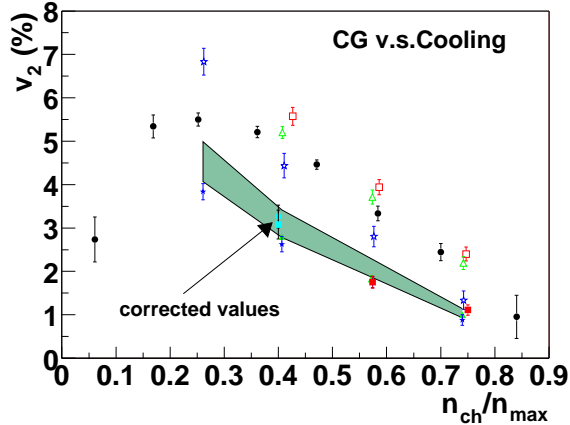} 
\end{center}
            \caption{The centrality dependence of 
            $v_2$ at early times 
            from cooling (open symbols) and CG 
            (filled symbols). The values of 
            $\Lambda_{s0}R$ span the RHIC-LHC range: 
            18.5 (squares), 37 (triangles), and 74 (stars). 
            Full circles are {\it preliminary} STAR data~\cite{Snellings}. 
            The band shows the range of $v_2$ extrapolated to 
            late times. ``Corrected values'' denote the late time cooling 
            and CG 
            result for $\Lambda_{s0}R=18.5$ at one centrality value. 
            } 
            \label{fig:v2CentDep} 
            \end{figure}

            The cooling and the CG results should converge at 
            late times, when the system is weakly 
            coupled. The two methods agree for $N$ at fairly early times. For 
            $v_2$, this 
            convergence occurs at much later $\tau$, because, as explained 
            below, $v_2$ is dominated by soft modes 
            with momenta $p_{\rm T}<\Lambda_{s0}$. Following the evolution of
the system to very late $\tau$ is computationally taxing. We therefore only
compute $v_2$ at late $\tau$ for a selected value of
$\Lambda_{s0}R$ and centrality and extrapolate $v_2$ from the early to the
late $\tau$ for the remaining values.
The results, for different values of $\Lambda_{s0} R$, are shown in 
Fig.~\ref{fig:v2CentDep} as a function of $n_{\rm{ch}}/n_{\rm{tot}}$.
            Clearly, our
            asymptotic values of $v_2$ undershoot 
            the 
            data.  Nevertheless, we see that an observable amount of 
            $v_2$ is produced classically 
            in the pre-equilibrium stage of a collision.

\begin{figure} 
\begin{center}
            \includegraphics[width=3.0in]{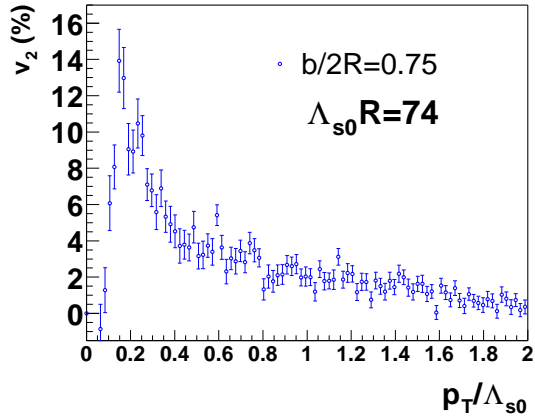} 
\end{center}
            \caption{Differential $v_2$ as a function of $p_{\rm T}$ in units of
$\Lambda_{s0}$ for
             $\Lambda_{s0}R=74$. 
            } 
            \label{fig:v2dndpt512} 
            \end{figure}

Our differential $v_2$, shown in Fig.~\ref{fig:v2dndpt512} for $b/2R=0.75$
and $\Lambda_{s0}R=74$, grows rapidly and is peaked for $p_{\rm{T}}\sim
            \Lambda_{s0}/4$.  A related analytical result
            ~\cite{TV} is that for $p_{\rm{T}} \gg \Lambda_{s0}$, 
            $v_2(p_{\rm{T}})\sim \Lambda_{s0}^2/p_{\rm{T}}^2$,  consistent with
our numerical data.  
The dominance of $v_2$ by very 
            soft modes helps explain the persistent difference between
the cooling and the CG values:
these modes remain strongly coupled
and cannot be described within a free theory
until 
            very late $\tau$. Concomitantly, the soft modes contain many gluons 
            and may be described classically  
            even at the late $\tau$ considered. 
            Our $v_2(p_{\rm{T}})$
            clearly disagrees with 
            experiment~\cite{STARv2}. 

            Note that experimental $v_2$ is found indirectly,
            in particular, from multiparticle 
            cumulants~\cite{Nicola}. It has been argued recently that 
non-flow
            correlations explain much of the measured 
$v_2$~\cite{KT}. We plan a numerical study of non-flow effects.

            We thank K. Filiminov, U. Heinz, D. Kharzeev, R. Lacey, Z. Lin, 
            L. McLerran, A. Mueller, J.-Y. Ollitrault, K. Rajagopal, J. Rak, 
            S. Voloshin, Nu Xu, E. Shuryak and D. Teaney for comments; 
            Sourendu Gupta for contributing to early stages of
            this work. We acknowledge support from: DOE Contract 
            No. DE-AC02-98CH10886 (R.V.), 
            the Portuguese FCT under grants 
            CERN/P/FIS/40108/2000 and CERN/FIS/43717/2001 (A.K., R.V.),
            RIKEN-BNL (R.V., Y.N.), and NSF Grant No. PHY99-07949. 
            A.K. thanks the BNL NTG  for hospitality.

            \end{document}